\documentclass[preprint,12pt]{elsarticle}

\usepackage{amssymb}

\usepackage{textcomp,booktabs}
\usepackage[usenames,dvipsnames]{color}
\usepackage{colortbl}
\definecolor{mygray}{gray}{.9}
\definecolor{mypink}{rgb}{.99,.91,.95}
\definecolor{mycyan}{cmyk}{.3,0,0,0}
\usepackage{amssymb}
\usepackage{graphicx}
\usepackage{booktabs}
\usepackage{tabularx}
\usepackage{array}
\usepackage{lscape}
\usepackage{longtable}
\usepackage{multirow}
\usepackage{amsmath}
\usepackage{colortbl}
\usepackage{epsfig}
\usepackage{epsf}

\usepackage{subfig}
\usepackage{paralist}
\usepackage{mdwlist}
\usepackage{calc}

\usepackage{algorithm}
\usepackage{algorithmicx}
\usepackage{algpseudocode}

\newcommand{\PreserveBackslash}[1]{\let\temp=\\#1\let\\=\temp}
\newcolumntype{C}[1]{>{\PreserveBackslash\centering}p{#1}}
\newcolumntype{R}[1]{>{\PreserveBackslash\raggedleft}p{#1}}
\newcolumntype{L}[1]{>{\PreserveBackslash\raggedright}p{#1}}

\journal{arXiv.org}

\begin{document}
\begin{frontmatter}



\title{A modified Physarum-inspired model for the user equilibrium traffic assignment problem}

\author[address1]{Shuai Xu}
\author[address1]{Wen Jiang\corref{label1}}
\author[address1]{Yehang Shou}
\address[address1]{School of Electronics and Information, Northwestern Polytechnical University, Xi'an, Shannxi, 710072, China}
\cortext[label1]{Corresponding author: School of Electronics and Information, Northwestern Polytechnical University, Xi'an, Shannxi, 710072, China. Tel:+86 029 88431267; fax:+86 029 88431267. E-mail address: jiangwen@nwpu.edu.cn, jiangwenpaper@hotmail.com}

\begin{abstract}
The user equilibrium traffic assignment principle is very important in the traffic assignment problem. Mathematical programming models are designed to solve the user equilibrium  problem in traditional algorithms.
Recently, the \emph{Physarum} shows the ability to address the user equilibrium and system optimization traffic assignment problems. However, the \emph{Physarum} model are not efficient  in real traffic networks with two-way traffic characteristics and multiple origin-destination pairs. In this article, a modified \emph{Physarum}-inspired model for the user equilibrium problem is proposed.
By decomposing traffic flux based on origin nodes, the traffic flux from different origin-destination pairs can be distinguished in the proposed model. The \emph{Physarum} can obtain the equilibrium traffic flux when no shorter path can be discovered between  each origin-destination pair. Finally, numerical examples use the Sioux Falls network to demonstrate  the rationality and convergence properties of the proposed model.
\end{abstract}


\end{frontmatter}


\section{INTRODUCTION}\label{Introduction}

The traffic assignment refers to a manner in which a given aggregate origin-destination (OD) pair passenger traffic demand is assigned to the traffic routes of that OD pair\cite{bertsekas1982projection, papageorgiou1990dynamic, ISI:000085665800003, ISI:000283918200005, du2016analysis}.  As an important traffic assignment paradigm, the user equilibrium (UE) of travelers' path choice in traffic networks was firstly conceptualised by Wardrop \cite{Wardrop1952Some}.
The UE principle is based on the assuming that the traveller knows the precise route cost and will choose the route with the minimum cost.  The user equilibrium is achieved when  all travellers between the same OD pair have the same and  minimum cost \cite{Beckmann1956Studies, Chiou2009An}. Additionally, the transportation cost of any traveller can't be reduced by unilaterally changing routes.

In the existing literatures, many algorithms were designed to address the UE problem \cite{florian2009new, Chiou2010An, Lin2014An}, which can be totally classed into three types: link based algorithms, path based algorithms and  origin based algorithms.
 The linear approximation method of Frank and Wolfe (FW) \cite{Leblanc1975An} has been the most popular algorithm
 in practice because of its  simplicity. As a linear approximation method, the FW algorithm has a sublinear asymptotic convergence speed \cite{florian2009new}. As a  consequence, highly precise solutions can't be achieved  within reasonable computation time.
In path based algorithms, the UE problem is solved by achieving the path flux directly \cite{ryu2014modified}. The traffic flux is assigned by fixing the flux from other OD pairs for each OD pair. In the exiting path based algorithms, the  disaggregate simplicial decomposition  algorithm (DSD) \cite{larsson1992simplicial} and the gradient projection algorithm (GP) \cite{jayakrishnan1994faster, cheng2003alternative, florian2009new} are widely used. Both DSD and GP have shown excellent results when compared with  FW algorithm \cite{sun1996computational, tatineni1998comparison}, but the  memory requirement is usually regarded as impractical for large-scale networks.
The origin based algorithm proposed by Bar-Gera \cite{bar1999origin, bar2002origin} doesn't require as much memory as  the path based algorithms.
The algorithm computes sequentially for each origin subnetwork by using a quasi-Newton approach. Later, Dial \cite{dial2006path} developed a different origin based algorithm by  shifting flows sequentially from the longest to the shortest path. Though the origin based algorithms can provide both link traffic flows and route traffic flows \cite{Xu2008An}, enumerating all route flows of each subnetwork is quite complicated.

Biological systems  usually inspire computer scientists and engineers to process the information  and make decisions. So far, some heuristic algorithms have been proposed to solve the  traffic assignment problem, such as the  ant colony algorithm \cite{Matteucci2013An, d2012ant} and the genetic algorithm \cite{MorenoDiaz2012}. Recently, the slime mould \emph{Physarum} polycephalum becomes a popular living computing substrate \cite{adamatzky2007physarum, adamatzky2010programmable}.
\emph{Physarum} machines are proved to be the most successful biological substrates in solving problems of computation geometry, optimization, and logic because they are easy to realize \cite{adamatzky2015using}.
In this article, a modified \emph{Physarum}-inspired model is proposed to solve the UE traffic assignment problem.

\emph{Physarum} polycephalum is a large amoeboid organism, which contains a great number of nuclei and tubular structures \cite{Stephenson1995Myxomycetes}. These tubular structures will distribute protoplasm as a transportation network.
 The experiment has shown that the \emph{Physarum} has the capacity of finding the short path between two points in a given labyrinth \cite{nakagaki2000intelligence}.
 A mathematical model can capture the basic dynamics of network adaptability through iterations of local rules and produces solutions with properties comparable to or better than those of real-world infrastructure networks¡¡\cite{Tero2010Rules}. The convergence of  \emph{Physarum} finding the shortest path has been proved by Bonifaci \emph{et al.} \cite{Bonifaci2012Physarum}. Based on its foraging¡¡behavior, so far, the \emph{Physarum} has been used to solve many problems, such as  finding the shortest path in directed or undirected network \cite{Adamatzky2012Slime, Wang2014A, Zhang2014An, Wang2015An},
designing and simulating transport network \cite{Adamatzky2012Bioevaluation, ANDREW2013BIO, tsompanas2014physarum, Evangelidis2015Slime},
natural implementation of spatial logic \cite{schumann2011logical, schumann2011physarum}, and computer music \cite{braund2013music, miranda2014harnessing, braund2015music}. In addition, the \emph{Physarum} model can also find the  shortest path under uncertain environment \cite{Zhangya2013fuzzy,Zhang2014fuzzy} in the real application \cite{Jiang2015improved}.

Recently, quite a few researchers try to apply the \emph{Physarum} model to solve the traffic assignment problem \cite{Zhang2015A, Zhang2015AnEfficient}.  These \emph{Physarum}-inspired  models can affect  well in specific conditions where networks are unilaterally connecting. But the exiting models  can't be  realised in traffic networks with two-way traffic characteristics.
The \emph{Physarum} model for UE traffic assignment problem \cite{Zhang2015AnEfficient}  can't distinguish the flux from different  OD pairs, which means the model isn't reasonable  in traffic networks with multiple OD pairs.
In this paper, a modified  \emph{Physarum}-inspired model for the UE traffic assignment problem is proposed for traffic networks with two-way traffic characteristics. In the proposed model, the flows are decomposed by origin node as the origin based algorithms. For each subnetwork, the flows are assigned by the \emph{Physarum} model based on its protoplasmic network adaptivity and  continuity.

 This paper is structured as follows. In Section 2, the user equilibrium traffic assignment principle is reviewed. The original \emph{Physarum} polycephalum model and \emph{ Physarum}-inspired model for UE traffic assignment problem proposed by
 Zhang \cite{Zhang2015AnEfficient} are briefly introduced. In Section 3,  a modified \emph{Physarum}-inspired model for the UE traffic assignment  problem is presented. In Section 4, numerical examples are given to  demonstrate the rationality and convergence properties of the proposed model. Finally, the paper ends with conclusions and suggestions for further researches  in Section 5.

\section{PRELIMINARIES}\label{PRELIMINARIES}

In this section, some preliminaries are briefly introduced, including the traffic assignment model \cite{Beckmann1956Studies}, the original \emph{Physarum} polycephalum model \cite{tero2007mathematical} and  the \emph{Physarum}-inspired model for UE problem proposed by Zhang \cite{Zhang2015AnEfficient}.

\subsection{User Equilibrium Traffic Assignment Model}

The transportation network is a strongly connected directed graph  ${G(N,A)}$, where $N$ is the set of nodes and $A$ is the set of directed links.
Assume that there is no links from a node to itself and only one link, if any, between two different nodes. Suppose that $R$  and $S$ denote the set of origin nodes and the set of destination nodes, and $r \in R$, $s \in S$, $R, S \subseteqq  N$.
 Let $K_{rs}$ and  $q_{rs}$ denote the set of all the paths and  the travel demand between OD pair $r-s$, than we have:
 \begin{equation}\label{fkrs}
 \sum\limits_{k \in {K_{rs}}} {f_k^{rs}}  = {q_{rs}}, ~~~{\rm{    }}\forall r \in R,s \in S
\end{equation}
where $f_k^{rs}$ is the path traffic flow along path $k$ between OD pair $r-s$.
Let $x_a$ denote the traffic flow along the link $a$.  Then all nodes, except source nodes and  destination nodes, satisfy the flow conservation law \cite{Zhang2015AnEfficient}, shown as follows:
\begin{equation}\label{xa}
\sum\limits_{a \equiv [(i,j) \in A]} {{x_a}}  = \sum\limits_{b \equiv [(j,k) \in A]} {{x_b}} ,\forall j,k \in N\backslash \{ R,S\}
\end{equation}

Let $t_{a}(x_a)$ denote the traffic time experienced by each user among the link $a$ when $x_a$ units of vehicles flux along the link.
$t_{a}(.)$ is a  monotonously non decreasing and continuously differentiable traffic time functions for the flux on link $a$ due to the effects of congestion on the travel time \cite{Beckmann1956Studies}, which can be expressed as follows:
\begin{equation}\label{taxa}
~~~~~\frac{{\partial {t_a}({x_a})}}{{\partial {t_b}}} = 0, ~~~ \forall a \ne b
\end{equation}
\begin{equation}\label{taxa}
\frac{{\partial {t_a}({x_a})}}{{\partial {x_a}}} > 0, ~~~ \forall a
\end{equation}

Let $c_k^{rs}$ represent the path traffic time along the path $k$ between OD pair $r-s$ and $\delta _{a,k}^{rs}$ denote the correlation coefficient between traffic link and traffic path, $\delta _{a,k}^{rs}=1$ if the $k_{th}$ path between OD pair $r-s$ traverses link $a$, otherwise $\delta _{a,k}^{rs}=0$. The path traffic time  $c_k^{rs}$ and link traffic flow $x_a$ can be expressed as \cite{Beckmann1956Studies}:
\begin{equation}\label{ckrs}
c_k^{rs} = \sum\limits_a^{} {{t_a}}  \cdot \delta _{a,k}^{rs} ~~~{\rm{    }}\forall r,s,k
\end{equation}
\begin{equation}\label{xaall}
{x_a} = \sum\limits_r {\sum\limits_s {\sum\limits_k {f_k^{rs} \cdot \delta _{a,k}^{rs}} } } , ~~~\forall a
\end{equation}

The Wardrop's user equilibrium principle \cite{Wardrop1952Some} is that travelers seek to minimize the cost associated with their chosen routes.
Travelers are assumed to have perfect information about actual travel conditions, and they can be identical in the sense that they valued time, monetary cost, and other route attributes in the same way. The Wardrop's user equilibrium is obtained when no traveller¡¯s traffic time  can be reduced by unilaterally changing routes, which can be expressed as follows:
\begin{equation}\label{}
{u_{rs}} - c_k^{rs}\left\{ \begin{array}{l}
  = 0,~~~f_k^{rs} > 0 \\
  \le 0,~~~f_k^{rs}{\rm{ = }}0 \\
 \end{array} \right.,~~~ \forall k,r,s
\end{equation}
where $u_{rs}$ is the shortest traffic time  between OD pair $r-s$ under traffic equilibrium.

Under the assumptions above all, the traffic assignment problem can be mathematically formulated as  the following convex optimization problem \cite{Beckmann1956Studies}:
\begin{equation}\label{optimization}
\begin{array}{l}
 \min Z(x) = \sum\limits_a {\int_0^{{x_a}} {{t_a}(w)dw} }  \\
 s.t.\\
 ~~~~~~{x_a} = \sum\limits_r {\sum\limits_s {\sum\limits_k {f_k^{rs} \cdot \delta _{a,k}^{rs}} } } , ~~~\forall a\\
 ~~~\sum\limits_{ [(i,j) \in A]} {{x_a}}  = \sum\limits_{ [(j,k) \in A]} {{x_b}} ,~~~~~~~\forall j,k \in N\backslash \{ R,S\}\\
 ~~~~~~\sum\limits_k {f_k^{rs}}  = {q_{rs}}, ~~~~~~~~~~~~~~~~\forall r,s\\
 ~~~~~~f_k^{rs} \ge 0, ~~~~~~~~~~~~~~~~~~~~~~\forall k,r,s\\
 ~~~~~~x_a \ge 0,~~~~~~~~~~~~~~~~~~~~~~~\forall a
 \end{array}
\end{equation}

\subsection{The Origin Physarum  Polycephalum Model }
\emph{Physarum} polycephalum is a single-celled amoeboid organism, which is also called as plasmodium in the vegetative phase \cite{tero2007mathematical}. It has the ability to solve the shortest path selection, based on its special foraging mechanism: the transformations of tubular structures and a positive feedback from flow rates. The high rates of the flow motivate tubes to thicken and the diameter of the tube minishes at a low flow rate \cite{tero2007mathematical}. A total introduction for the \emph{physarum}
polycephalum  path finding model is given below.

Suppose the shape of the network formed by the Physarum is represented by a graph $G(N,A)$, where the edge of the graph denotes the plasmodial tube and the node denotes the junction between tubes.
And $N$ is a set of nodes, where $N_1$ and $N_2$ are
signed as the source and destination nodes, any others are labeled as $N_3$, $N_4$, $N_5$, etc. The edge connecting node $i$ and node $j$ are remarked as $M_{ij}$ and the flux from node $N_i$ to node $N_j$ through edge $M_{ij}$ is remarked as ${Q_{ij}}$, shown as follows \cite{tero2007mathematical}:
\begin{equation}\label{Qij}
{Q_{ij}} = \frac{{\pi {\rm{r}}_{ij}^4}}{{8\eta {L_{ij}}}}({p_i} - {p_j}) = \frac{{{D_{ij}}}}{{{L_{ij}}}}({p_i} - {p_j})
\end{equation}
where $\eta$ is the viscosity of the fluid and ${D_{ij}}{\rm{ = }}\pi {\rm{r}}_{ij}^4/8\eta$ is the measure of the conductivity of the edge tube $M_{ij}$.
And $p_i$ is the measure of the pressure at the node $n_i$, $L_{ij}$ is the length of the edge  $M_{ij}$.
According to the conservation law of flow, the inflow and outflow must be balanced, namely:
\begin{equation}\label{Q}
\sum {{Q_{ij}} = } 0,~~~{\rm{    }}({\rm{j }} \ne {\rm{1,2}})
\end{equation}
especially for the source nodes $N_1$ and $N_2$, the flux equations can be expressed as:
\begin{equation}\label{Q1}
\sum\limits_i {{Q_{i1}}}  + {I_0} = 0
\end{equation}
\begin{equation}\label{Q2}
\sum\limits_i {{Q_{i2}}}  - {I_0} = 0
\end{equation}
where $I_0$ is the flux from the source node to the destination node, which is assumed as a constant in the model. According to the Eqs.(\ref{Qij}-\ref{Q2}), the network Poisson equation for the pressure is derived as following:
\begin{equation}\label{Dij}
\sum\limits_i {\frac{{{D_{ij}}}}{{{L_{ij}}}}({p_i} - {p_j})}  = \left\{ \begin{array}{l}
  - 1~~~~~{\rm{  }}for~{\rm{}}j = 1, \\
  + 1~~~~~{\rm{  }}for~{\rm{}}j = 2, \\
    0~~~~~~~{\rm{    }}otherwise~~~{\rm{}} \\
 \end{array} \right.
\end{equation}
by further setting $p_2=0$ as the basic pressure level, the pressure of all nodes can be determined, the pressure of all nodes can be determined  according to Eq.(\ref{Dij}) and all $Q_{ij}$ can  be determined by solving Eq.(\ref{Qij}).

To accommodate the adaptive behavior of the plasmodium, the conductivity $D_{ij}$ is assumed to change when adapting to the flux $Q_{ij}$. And tubes with zero conductivity will die out. The conductivity of each tube is described as follows: \cite{Tero2010Rules}:
\begin{equation}\label{Dt}
\frac{d}{{dt}}{D_{ij}} = f(|{Q_{ij}}|) - \alpha {D_{ij}}
\end{equation}
where $\alpha$ is the decay rate of the tube and $f(.)$ is monotonously increasing continuous function which satisfies $f(0) = 0$. Obviously, the positive feedback exists in the model.

\subsection{The Physarum-inspired Model for UE Problem}\label{Physarum-inspired}

According to the feature of \emph{Physarum} foraging behavior system, the optimum problem and the user equilibrium problem in directed traffic networks are solved by  modified \emph{Physarum} models \cite{Zhang2015A, Zhang2015AnEfficient} . There are mainly two points different from from the original \emph{Physarum} model  in Zhang's method \cite{Zhang2015AnEfficient}.

\begin{figure}[htb]

\subfloat[The origin model]{%
  \includegraphics[width=.48\textwidth]{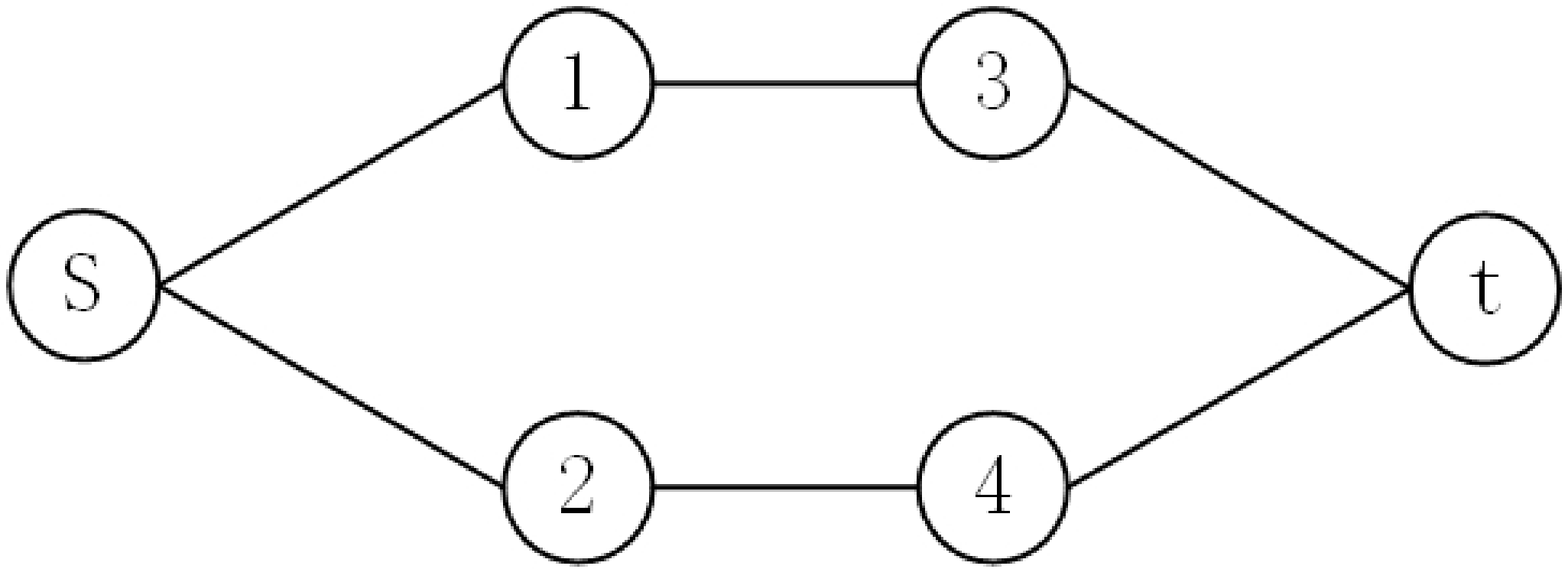}}\hfill
\subfloat[The model modified by Zhang \cite{Zhang2015AnEfficient}]{%
  \includegraphics[width=.48\textwidth]{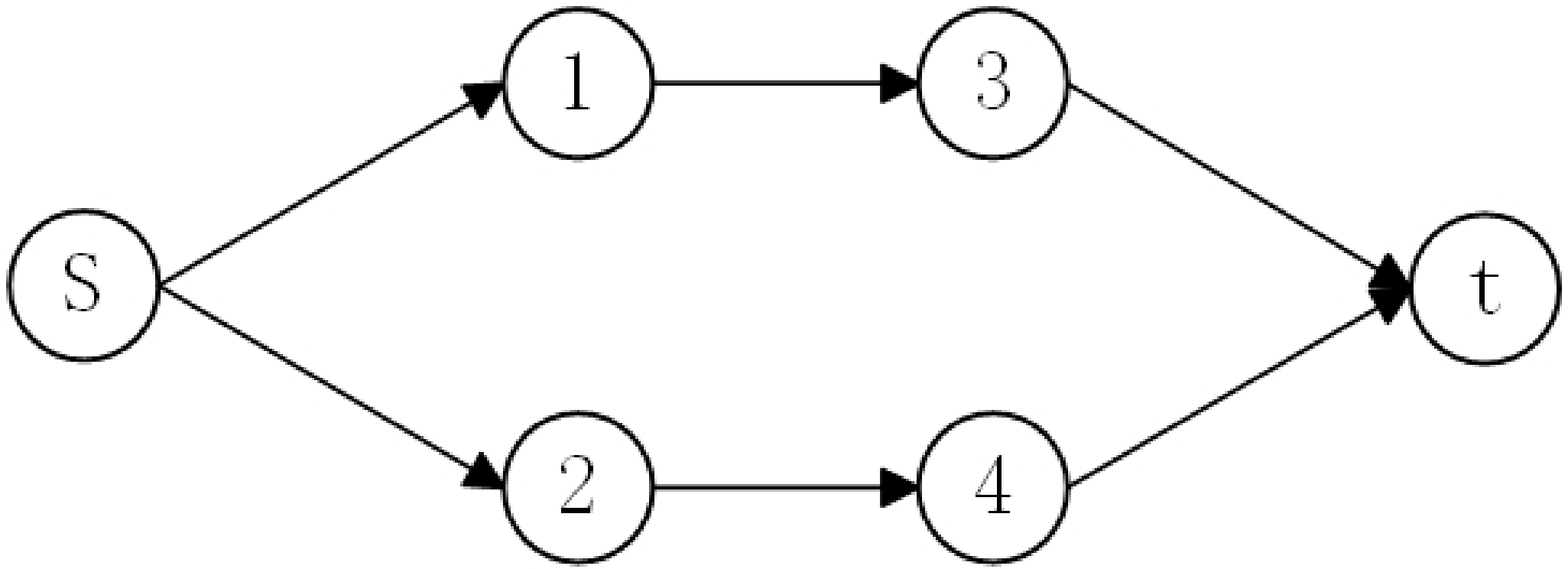}}
\caption{Simple networks for \emph{Physarum} model}
\label{fig1}
\end{figure}

First, the original \emph{Physarum} model can only find the shortest path in undirected networks shown in Figure \ref{fig1}(a). The modified method was proposed to extend the original \emph{Physarum} model to directed networks. A total introduction for the modified model is given below.

In the modified model, each edge is regard as two tubes with opposite directions and equal weight, which is shown in Figure \ref{fig1}(b).  And there is only one direction between two nodes in the network, which means that flux can only flow from node $s$ to node $1$ and it can't flow in the opposite direction as shown in Figure \ref{fig1}(b). And the network Posson equation defined in Eq.(\ref{Dij}) was modified as follows \cite{Zhang2015AnEfficient}:
\begin{equation}\label{modifiedDij}
\sum\limits_i {(\frac{{{D_{ij}}}}{{{L_{ij}}}} + \frac{{{D_{ji}}}}{{{L_{ji}}}}) + ({p_i} - {p_j})}  = \left\{ \begin{array}{l}
  - 1~~~~~{\rm{  }}for~{\rm{}}j = 1, \\
  + 1~~~~~{\rm{  }}for~{\rm{}}j = 2, \\
    0~~~~~~~{\rm{    }}otherwise~~~{\rm{}} \\
 \end{array} \right.
\end{equation}
 where $L_{ij}$ denotes the travel time of from node $i$ to node $j$ and $L_{ji}$ denotes the travel time of from node $j$ to node $i$. Similarly, $D_{ij}$ and $D_{ji}$ have different meanings. In the initialization, if $L_{ij}=inf$, then $D_{ij}=0$. Otherwise, assign a value between $0$ and $1$ to $D_{ij}$. To guarantee the inconsistency, $Q_{ij}$ should be kept $0$ during  iterations if $L_{ij}=inf$.

 Second, there is only one source and one destination in networks. However, usually there are multiple OD pairs in traffic networks. To handel the network with multiple origins and destinations, Eq.(\ref{modifiedDij}) was modified as follows \cite{Zhang2015AnEfficient}:
 \begin{equation}\label{multiDij}
\sum\limits_i {(\frac{{{D_{ij}}}}{{{L_{ij}}}} + \frac{{{D_{ji}}}}{{{L_{ji}}}})({p_i} - {p_j})}  = \left\{ \begin{array}{l}
  - {I_r},~~~~~{\rm{  }}\forall r \in R, \\
  + {I_s},~~~~~{\rm{  }}\forall s \in s, \\
 0,~~~~~~~~~{\rm{  }}otherwise \\
 \end{array} \right.
 \end{equation}
where $r$ denotes the origin node and $r \in  R$, $s$ denotes the destination node and $s \in S$. $I_r$ and $I_s$ represent the units of flow supplied by the origin node $r$ and consumed by the destination node $s$.

 \section{PROPOSED METHOD}\label{Proposed method}

In this section, we will discuss the shortcomings of the original \emph{Physarum}-inspired method for UE problem and propose the modified \emph{Physarum}-inspired method to solve the UE problem.

\subsection{Shortcomings of the Origin Physarum-inspired Method}\label{shortcomings}

Compared with the previous slime mould models, the \emph{Physarum}-inspired method for UE problem is superior  when characterizing its foraging activity. However, two shortcomings of the original \emph{Physarum}-inspired method for UE problem are founded as follows:
\begin{enumerate}
\item Note that there is only one direction between two nodes in the network shown in Figure \ref{fig1}(b), which means the flux can only flow from one node to another but never in the opposite direction. However, in real traffic networks, most roads have the properties of two-way traffic characteristics as shown in Figure \ref{fig2}. Clearly, opposite directions are separated with each other, where  flows don't interfere in two opposite directions. Obviously, the original \emph{Physarum}-inspired method can't be implemented in the traffic network shown in Figure \ref{fig2}.
    \begin{figure}[!ht]
\centering
\includegraphics[scale=0.4]{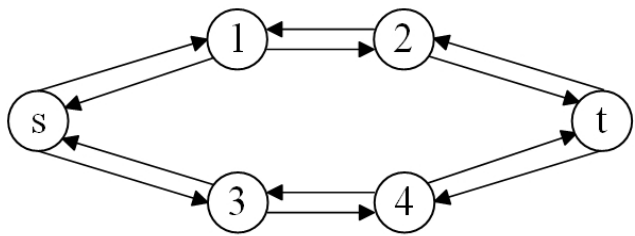}
 \caption{The real traffic network} \label{fig2}
\end{figure}
\item The \emph{Physarum}-inspired model for UE problem isn't reasonable in traffic networks with  multiple OD pairs. The modified equation Eq.(\ref{multiDij}) only satisfies the flow conservation law of the traffic network. However, it can't distinguish the flux in each OD pair, which means that for a given OD pair $r-s$, the output flux at node $s$ doesn't equal the input flow at node $r$. In fact, the \emph{Physarum}-inspired model for UE problem is only suitable for traffic networks with one source and multiple sinks or multiple sources and one sink. Here,  we illustrate this problem with the following example.
\end{enumerate}
\noindent \emph{Example $~$ A small network with multiple sources and multiple sinks}

\begin{figure}[!ht]
\centering
\includegraphics[scale=0.4]{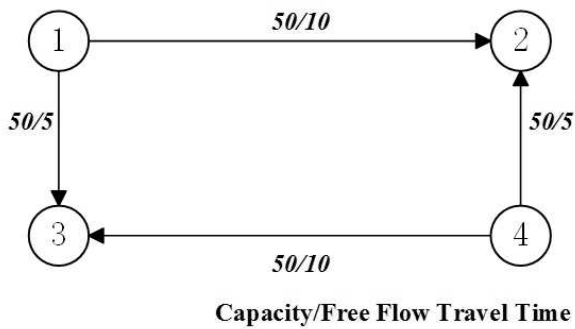}
 \caption{A small traffic network} \label{Example1}
\end{figure}

\textit{
Here, we use a small traffic network with 4 nodes, 4 links and 2 OD pairs which are shown in Figure \ref{Example1}. And the origin-destination demands, in vehicles per hour, are $q_{1,2}$=100 and $q_{4,3}=100$.
For simplicity, the link travel time is calculated by the US Bureau of Public Roads(BPR) function \cite{hansen1959accessibility}, which is expressed as follows:
\begin{equation}\label{BPR}
{t_a} = {\alpha _a}(1 + 0.15{(\frac{{{Q_a}}}{{{c_a}}})^4})
\end{equation}
where $t_a$, $\alpha _a$, $Q_a$, $c_a$ denote the travel time, free flow travel time flow and the capacity on link $a$, respectively.}

\textit{
By using the \emph{Physarum}-inspired model, the path flows are shown in Table \ref{tableExa}. It's clear that traffic assignment doesn't satisfy the traffic demands, which the flux along path $1 \rightarrow 3$  and path $4 \rightarrow 2$  should equal 100.  Note that the free flow travel time of $1 \rightarrow 3$ and $4 \rightarrow 2$ is much less than that of $1 \rightarrow 2$ and $4 \rightarrow 3$. Clearly the model doesn't distinguish the flux from different OD pairs. As a result, the \emph{Physarum}-inspired model assigns more flux to path $1 \rightarrow 3$ and $4 \rightarrow 2$.}

\begin{table}[!h]
{\scriptsize%
\caption{Path flows of the traffic network shown in Figure \ref{Example1} }\label{tableExa}
\begin{tabular*}{\columnwidth}{@{\extracolsep{\fill}}@{~~}ccc@{~~}}
\toprule
  Path(node sequence)  & Path flow & Path cost \\
\midrule
  1 $\rightarrow$ 2  &19.5177  & 10.0348   \\
  1 $\rightarrow$ 3 & 80.4823 &10.0348 \\
  4 $\rightarrow$ 2 & 80.4823 & 10.0348\\
  4 $\rightarrow$ 3  & 19.5177  & 10.0348 \\
\bottomrule
\end{tabular*}
}
\end{table}

\subsection{The Modified Physarum-inspired Model for UE Problem }\label {modified physarum}
Here, we  proposed a modified physarum-inspired model for UE problem to overcome these  above-mentioned shortcomings.

\subsubsection{Modified Physarum model for the shortest path in directed networks}

To satisfy the characteristic of the real traffic network shown in Figure \ref{fig2},  each edge is regarded as two tubes with opposite directions and different weight. And flux can flow in opposite directions without interfering with each other. The network Posson equation is same as Eq.(\ref{modifiedDij}). But to maintain the validity of conductivity, the conductivity equation defined
in Eq.(\ref{Qij}) should be improved as follows:
\begin{equation}\label{modiQIJ}
{Q_{ij}} = \left\{ \begin{array}{l}
 \frac{{{D_{ij}}}}{{{L_{ij}}}}({p_i} - {p_j}),~~~~~{\rm{  }} \frac{{{D_{ij}}}}{{{L_{ij}}}}({p_i} - {p_j}) > 0 \\
0,~~~~~~~~~~~~~~~~~~~{\rm{    }}otherwise~~~{\rm{}} \\
 \end{array} \right.
\end{equation}
when the flux in each tube is negative, the flux will be assigned as 0. That's because the flux can't get through the given tube in the opposite direction. In the initialization, if ${L_{ij} \ne inf}$, we assign a value between 0 and 1 to ${D_{ij}}$. Otherwise, conductivity ${D_{ij}}$ is assigned as $0$.

Particularly, when $L_{ij}= inf$ or $L_{ji}= inf$, which means that the flux can only flow from node $j$ to node $i$ or from node $i$ to node $j$, our modified model is the same as that introduced in Sec. \ref{Physarum-inspired}. Exactly, the model shown in Sec. \ref{Physarum-inspired} is a special case of our modified model.

\subsubsection{Modified Physarum model for directed networks with sources and multiple sinks}\label{sec322}

To overcome the defect that the \emph{Physarum}-inspired model can't  distinguish the flux in each OD pair, we modify the \emph{Physarum}-inspired model as below. Let $S_r$ denote the set of destination nodes which are originated from node $r$. Clearly, we have:
\begin{equation}\label{Ir}
{I_r} = \sum\limits_{s \in S_r} {{I_{rs}}}
\end{equation}
where $I_r$ denotes the input flow at the origin node $r$ and  $I_{rs}$ denotes the output flow at the destination node $s$ originated from node $r$. Here, we regard the multiple sources and multiple sinks network as the superposition of one source and multiple sinks networks. For each one source and multiple sinks network, we modify the original network Poisson equation Eq.(\ref{Dij}) as follows:
\begin{equation}\label{subnetworkDij}
\sum\limits_i {(\frac{{{D_{ij}}}}{{{L_{ij}}}} + \frac{{{D_{ji}}}}{{{L_{ji}}}})({p_i} - {p_j})}  = \left\{ \begin{array}{l}
  - {I_r},~~~~~{\rm{  }}for ~~j=r,\\
  + {I_{rs}},~~~~~{\rm{  }}\forall s \in S_{rs}, \\
 0,~~~~~~~~~{\rm{  }}otherwise \\
 \end{array} \right.
\end{equation}

Naturally, the flux of each tube in all one source and multiple sinks subnetworks can be calculated by Eq.(\ref{modiQIJ}). According to the superposition principle, the flux of each tube in the original multiple sources and sinks network can be expressed as follows:
\begin{equation}
Q_{ij}^{all} = \sum\limits_{r \in R} {Q_{ij}^r},~~~\forall i,j \in N
\end{equation}
where $Q_{ij}^{all}$ denotes the flux from node $i$ to node $j$ in the original multiple sources and multiple sinks network and $Q_{ij}^r$ represents the flux from node $i$ to node $j$  in the subnetwork originated from node $r$.

\subsubsection{Modified Physarum model for UE problem}
 Note that in the process of \emph{Phyasrun} approaching the shortest path, the flow and the conductivity along each link are continuous.  Further consideration of the continuity  and dynamic reconfiguration of \emph{Physarum} model, we can update the link travel time within each  iteration. The flux will be redistributed by the modified \emph{Physarum} model when the link travel time is updated during iterations. The length of link $a$ is updated as follows:
 \begin{equation}\label{}
 L_a^{n + 1} = \frac{{L_a^n + {t_a}(Q_a^{all_{n + 1}})}}{2}
 \end{equation}
 where $Q_a^{all_{n + 1}}$ denotes the total flow on link $a$ at the $(n+1)_{th}$ iteration, $L_a^{n}$ and $L_a^{n+1}$ represent the length of link $a$ at the $n_{th}$ and $(n+1)_{th}$ iteration. And the search direction of link length $L_a$ is guided by ${t_a}(Q_a^{all_{n + 1}})$. Note that in equilibrium, there will be $L_a=t_a(Q_a^{all})$, which means the length of link $a$ equals the travel time along link $a$.

Here, the main steps of the modified \emph{Physarum}-inspired model for user equilibrium problem is presented in Algorithm 1.

\begin{algorithm}
  \caption{Modified \emph{Physarum}-inspired model for UE problem}
  \begin{algorithmic}
   \State // $Q^r$ is the flow matrix of the subnetwork originated from node $r$.
   \State // $L^0$ is the free flow travel time matrix.
   \State $D_{ij}^r = [0.5,1](\forall i,j = 1,2,\cdots,N \wedge C_{ij}^0 \ne 0,\forall r \in R)$
   \If {$C_{ij}^0 == inf $}
        \State $D_{ij}^r = 0$
   \EndIf
   \State $Q_{ij}^r = 0(\forall i,j = 1,2,\cdots,N,\forall r \in R )$
    \State ${ Q_{ij}^{all_{1}}} = 0(\forall i,j = 1,2,\cdots,N )$
    \State $L_{ij}^1=L_{ij}^0 (\forall i,j = 1,2,\cdots,N )$
    \State $n=1$ //Iteration counter

    \While{$NotConvergence$}
    \For{$r \in R$}
              \State //Calculate the pressure of every node using Eq.(\ref{subnetworkDij})
              \State
              $\sum\limits_i {(\frac{{{D_{ij}^r}}}{{{L_{ij}}}} + \frac{{{D_{ji}^r}}}{{{L_{ji}}}})({p_i} - {p_j})}  = \left\{ \begin{array}{l}
              - {I_r},~~~~~{\rm{  }}for ~~j=r,\\
              + {I_{rs}},~~~~~{\rm{  }}\forall s \in S_{rs}, \\
              0,~~~~~~~~~{\rm{  }}otherwise \\
              \end{array} \right.$
             \State //Calculate the flux of every edge using Eq.(\ref{modiQIJ})
             \State ${Q_{ij}^r} = \left\{ \begin{array}{l}
         \frac{{{D_{ij}^r}}}{{{L_{ij}}}}({p_i} - {p_j}),~~~~~{\rm{  }} \frac{{{D_{ij}^r}}}{{{L_{ij}}}}({p_i} - {p_j}) > 0 \\
        0,~~~~~~~~~~~~~~~~~~~{\rm{    }}otherwise~~~{\rm{}} \\
         \end{array} \right.$
            \State ${\rm{D}}_{ij}^{r} = ({\rm{D}}_{ij}^r + {Q_{ij}^r})/2,~~~\forall i,j \in N$
      \EndFor
     \State $Q_{ij}^{all_{n+1}} = \sum\limits_{r \in R}{Q_{ij}^r},~~~\forall i,j \in N$

       \State $L_{ij}^{n + 1} = \frac{{L_{ij}^n + {t_a}(Q_{ij}^{all_{n + 1}})}}{2}$
      \State $n = n +1$
   \EndWhile
  \end{algorithmic}
\end{algorithm}

\subsection{Discussion}
Usually, there are three convergence measures for the traffic assignment \cite{bar2002origin, boyce2004convergence}, which are briefly introduced as follows:
\begin{enumerate}[{Principle} 1 :]
\item   According to  the error of  traffic flows or travel time calculated in two adjacent times to decide  whether the iteration stops \cite{Leblanc1975An}. The computing will stop only when the flows become stable in two adjacent iterations. The measure is simple and usually effective, which can get the flows satisfying the UE principle. Naturally the convergence principle can be expressed as follows:
    \begin{equation}
     \left| {Q_{ij}^{all_{n+1}}-Q_{ij}^{all_{n}}} \right|\leqslant \varepsilon
      ~or~ \left| {L_{ij}^{n+1}- L_{ij}^{n}}\right| \leqslant C ~~~\forall i,j\in N
    \end{equation}
\item  According to the error between the travel time path-based and link-based network, the relative gap (RGAP) and normalized gap (or excess average cost) are taken into consideration to measure the convergence \cite{florian2009new}. The measure of RGAP can be expressed as:
    \begin{equation}
    \small
    RGAP = \frac{{({\rm{total~travel~ time}} - {\rm{total~ travel~ time ~on~ shortest ~paths}})}}{{{\rm{total~ travel ~ time}}}}
    \end{equation}
\item The error between the maximum path travel time  and the minimum path travel time between the OD pair is also an evaluation index of convergence \cite{bar2002origin}:
    \begin{equation}
   {\varepsilon _{rs}} = \max \{ c_k^{rs}|k \in {K^{rs}},f_k^{rs} > 0\}  - \min \{ c_k^{rs}|k \in {K^{rs}}\}
    \end{equation}
    by giving a suitable value to $\varepsilon _{rs}$, the UE principle can be archived.
\end{enumerate}

Usually, Principle 1 is used in the link based method, such as the FW method. Principle 2 is usually used in the path-based method and Principle 3 is generally used in the origin-based algorithm. Note that the modified \emph{Physarum}-inspired model for UE problem is actually basing on the link travel time. During iterations, the shortest travel paths and the maximum path travel time aren't available. In order to reduce computing time, we choose the Principle 1 as the  convergence measure  in this article, which can be expressed as:
\begin{equation}
\sum\limits_{i,j\in N} {\left | Q_{ij}^{all_{n+1}}{\rm{ - }} Q_{ij}^{all_n}\right |}\leqslant\varepsilon_0,~~~\forall i,j \in N
\end{equation}

In the proposed \emph{Physarum}-inspired algorithm, solving a linear system of equations  is necessary for each origin based subnetwork.
Note that many investigations about paralleled Physarum model have been achieved \cite{adamatzky2008towards}, it's clear that the paralleled computing can be implemented for each origin based subnetwork.
Besides, the linear system of equations for each origin based subnetwork is very special and can be formulated as Laplacian system, which is solvable in $O(m \log n)$ \cite{koutis2010approaching}.
For the sake of simplicity, paralleled computing isn't implemented in the following. And the linear system of equations for each origin based subnetwork is solved in a general method with $O(n^3)$.

\section{NUMERICAL EXAMPLES}\label{numerical examples}

In this section, numberical examples are illustrated to demonstrate the rationality and convergence properties of the modified method. And the effect of the stopping criterion $\varepsilon_0$ is further discussed.

Computational examples reported in this article   are using Matlab on Intel(R) Core(TM) i5-5200U processor (2.2Ghz) with 8.00 GB of RAM under Windows Eight.

\subsection{Example 1}

Here, we use the simple network shown in Figure \ref{Example1} and the OD demands are same as those showed in Sec. \ref{shortcomings}. By utilizing the modified  \emph{Physarum}-inspired model, the path flows are shown in Table \ref{tableExa1}. It's clear that the modified \emph{Physarum}-inspired model has the ability of distinguishing the flux in different OD pairs.
\begin{table}[!h]
{\scriptsize%
\caption{Path flows resulted from the modified method }\label{tableExa1}
\begin{tabular*}{\columnwidth}{@{\extracolsep{\fill}}@{~~}cccc@{~~}}
\toprule
  stopping criterion($\varepsilon_{0}$)   &Path(node sequence)  & Path flow  & Iteration \\
\midrule
  0.01      &1 $\rightarrow$ 2  &100  &17   \\
        &4 $\rightarrow$ 3      &100    \\
\bottomrule
\end{tabular*}
}
\end{table}

\subsection{Example 2}

In this example, we test the modified  \emph{Physarum}-inspired model on the Sioux Falls network \cite{Leblanc1975An} which  is used in many publications for the traffic assignment problem.

OD flows are given in thousands of vehicles per day, with integer values up to 44  \cite{Leblanc1975An}. OD flows here are the values from the table multiplied by 100. They are therefore 0.1 of the original daily flows, and in that sense might be viewed as approximate hourly flows. The parameters in \cite{Leblanc1975An} are given as:
\begin{equation}
{t_{ij}} = a+b{Q_{ij}}^4
\end{equation}
where $a$ denotes the free flow travel time given here. The original parameter $b$ can be expressed as:
\begin{equation}
{b} = B \cdot \frac{{{\alpha}}}{{{c^{Power}}}}
\end{equation}
where $\alpha$ and $c$ denote the free flow traffic time and the capacity flow, respectively. $Power$ is set as $4$ and assume the ¡°traditional¡± BPR value of $B=0.15$, so we can get the same travel time equation as Eq.(\ref{BPR}).  And the free flow travel time($\alpha$) and the capacity flow($c$) are shown in Table \ref{tableExa2}.
\begin{table}[!h]
{\scriptsize
\caption{Capacity flow($c$) and free flow travel time($\alpha$) along links }\label{tableExa2}
\begin{tabular*}{\columnwidth}{@{\extracolsep{\fill}}@{}llllllllllll@{}}
\toprule
  ARC&$c(10^4)$&$\alpha$      &ARCS    &$c(10^4)$       & $\alpha$   &ARCS    &$c(10^4)$       & $\alpha$ &ARCS    &$c(10^4)$       & $\alpha$       \\
  \midrule
 $(1,2)  $    &2.5900  & 6&$(1,3)  $    &2.3403  & 4&$(2,1)  $    &2.5900  & 6&$(2,6)  $    &0.4958  & 5   \\
$(3,1)  $    &2.3403  & 4&$(3,4)  $    &1.7111  & 4&$(3,12) $    &2.3403  & 4&$(4,3)  $    &1.7111  & 4   \\
$(4,5)  $    &1.7783  & 2&$(4,11) $    &0.4909  & 6&$(5,4)  $    &1.7783  & 2&$(5,6)  $    &0.4948  & 4   \\
$(5,9)  $    &1.0000  & 5&$(6,2)  $    &0.4958  & 5&$(6,5)  $    &0.4948  & 4&$(6,8)  $    &0.4899  & 2   \\
$(7,8)  $    &0.7842  & 3&$(7,18) $    &2.3403  & 2&$(8,6)  $    &0.4899  & 2&$(8,7)  $    &0.7842  & 3   \\
$(8,9)  $    &0.5050  &10&$(8,16) $    &0.5046  & 5&$(9,5)  $    &1.0000  & 5&$(9,8)  $    &0.5050  &10   \\
$(9,10) $    &1.3916  & 3&$(10,9) $    &1.3916  & 3&$(10,11)$    &1.0000  & 5&$(10,15)$    &1.3512  & 6   \\
$(10,16)$    &0.4855  & 4&$(10,17)$    &0.4994  & 8&$(11,4 )$    &0.4909  & 6&$(11,10)$    &1.0000  & 5   \\
$(11,12)$    &0.4909  & 6&$(11,14)$    &0.4877  & 4&$(12,3 )$    &2.3403  & 4&$(12,11)$    &0.4909  & 6   \\
$(12,13)$    &2.5900  & 3&$(13,12)$    &2.5900  & 3&$(13,24)$    &0.5091  & 4&$(14,11)$    &0.4877  & 4   \\
$(14,15)$    &0.5128  & 5&$(14,23)$    &0.4925  & 4&$(15,10)$    &1.3512  & 6&$(15,14)$    &0.5128  & 5   \\
$(15,19)$    &1.4565  & 3&$(15,22)$    &0.9599  & 3&$(16,8 )$    &0.5046  & 5&$(16,10)$    &0.4855  & 4   \\
$(16,17)$    &0.5230  & 2&$(16,18)$    &1.9680  & 3&$(17,10)$    &0.4994  & 8&$(17,16)$    &0.5230  & 2   \\
$(17,19)$    &0.4824  & 2&$(18,7 )$    &2.3403  & 2&$(18,16)$    &1.9680  & 3&$(18,20)$    &2.3403  & 4   \\
$(19,15)$    &1.4565  & 3&$(19,17)$    &0.4824  & 2&$(19,20)$    &0.5003  & 4&$(20,18)$    &2.3403  & 4   \\
$(20,19)$    &0.5003  & 4&$(20,21)$    &0.5060  & 6&$(20,22)$    &0.5076  & 5&$(21,20)$    &0.5060  & 6   \\
$(21,22)$    &0.5230  & 2&$(21,24)$    &0.4885  & 3&$(22,15)$    &0.9599  & 3&$(22,20)$    &0.5076  & 5   \\
$(22,21)$    &0.5230  & 2&$(22,23)$    &0.5000  & 4&$(23,14)$    &0.4925  & 4&$(23,22)$    &0.5000  & 4   \\
$(23,24)$    &0.5079  & 2&$(24,13)$    &0.5091  & 4&$(24,21)$    &0.4885  & 3&$(24,23)$    &0.5079  & 2   \\

\bottomrule
\end{tabular*}
}
\end{table}

\begin{figure}[!ht]
\centering
\includegraphics[scale=0.8]{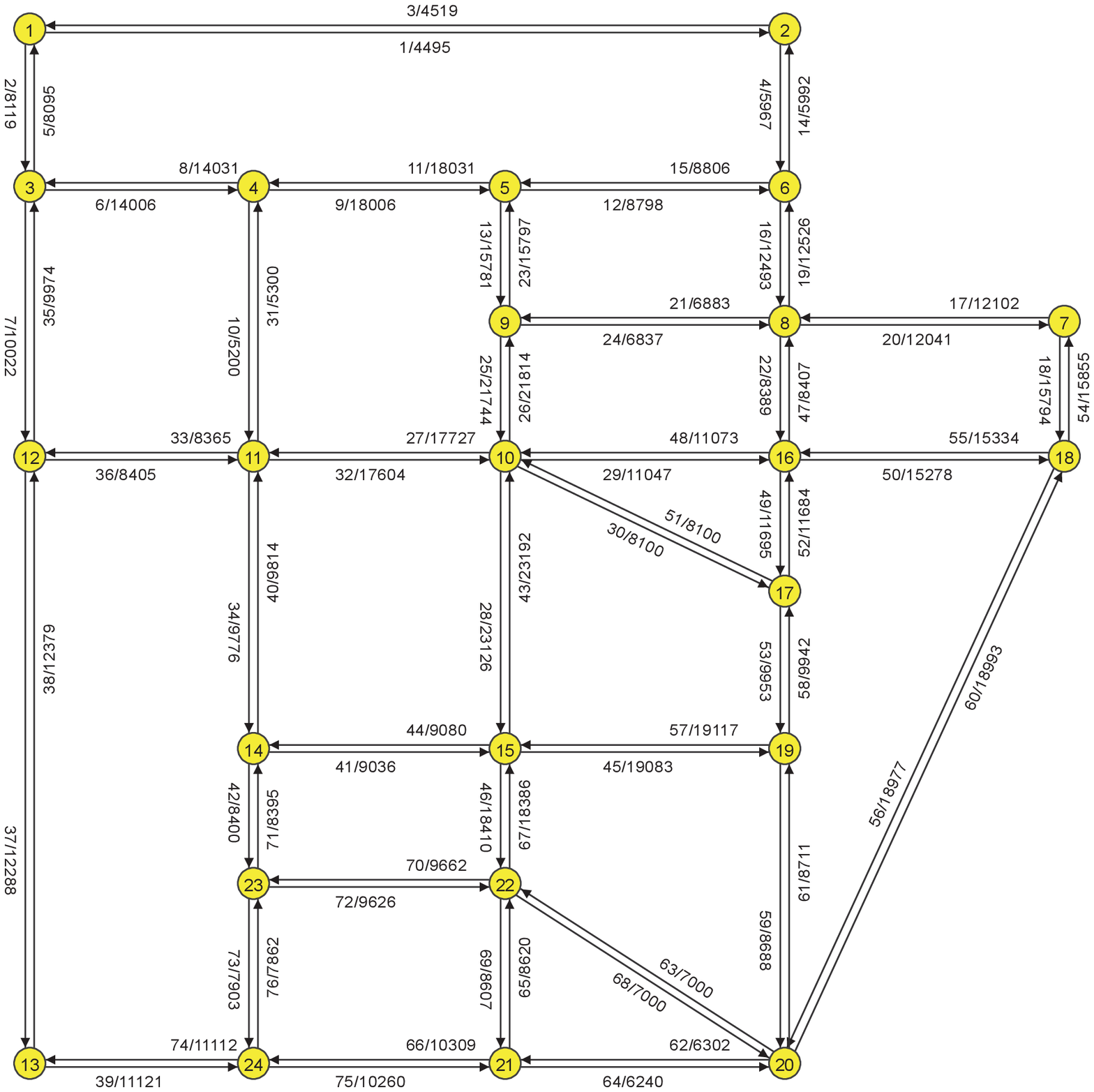}
 \caption{Equilibrium flows calculated by FW algorithm } \label{figFW}
\end{figure}

There are 76 arcs, 24 nodes, 552 conservations of flow constraints and 1824 nonnegativity constrains in the network.  Here, we set $\varepsilon=0.1$ as the stopping criterion of the modified \emph{Physarum}-inspired algorithm for the Sioux Falls network. And traffic flows of each link calculated by the proposed algorithm and FW algorithm  are shown in Table \ref{tableFlow} and Figure \ref{figFW}, respectively. Compared with the  FW algorithm, the proposed \emph{Physarum}-inspired algorithm obtains the same traffic flows.

\begin{table}[!h]

{
\scriptsize
\caption{The equilibrium flows calculated by the proposed algorithm }\label{tableFlow}
\begin{tabular*}{\columnwidth}{@{\extracolsep{\fill}}@{~~}llllllll@{~~}}
\toprule
  ARC   &$Q(10^3)$    &ARCS &$Q(10^3)$   &ARCS &$Q(10^3)$     &ARCS  &$Q(10^3)$         \\
  \midrule
$(1,2)  $    &    4.4945&$(1,3)  $    &    8.1189&$(2,1)  $    &    4.5189&$(2,6)  $    &    5.9674                      \\
$(3,1)  $    &    8.0945&$(3,4)  $    &   14.0068&$(3,12) $    &   10.0226&$(4,3)  $    &   14.0307                       \\
$(4,5)  $    &    18.0068&$(4,11) $    &     5.2000&$(5,4)  $    &    18.0307&$(5,6)  $    &     8.7983                   \\
$(5,9)  $    &    15.7812&$(6,2)  $    &     5.9919&$(6,5)  $    &     8.8065&$(6,8)  $    &    12.4928                    \\
$(7,8)  $    &    12.1012&$(7,18) $    &    15.7966&$(8,6)  $    &    12.5254&$(8,7)  $    &    12.0405                   \\
$(8,9)  $    &     6.8824&$(8,16) $    &     8.3886&$(9,5)  $    &    15.7969&$(9,8)  $    &     6.8363                   \\
$(9,10) $    &    21.7448&$(10,9) $    &    21.8145&$(10,11)$    &    17.7266&$(10,15)$    &    23.1267                   \\
$(10,16)$    &    11.0469&$(10,17)$    &     8.1000&$(11,4 )$    &     5.3000&$(11,10)$    &    17.6041                   \\
$(11,12)$    &     8.3654&$(11,14)$    &     9.7764&$(12,3 )$    &     9.9742&$(12,11)$    &     8.4052                   \\
$(12,13)$    &    12.2881&$(13,12)$    &    12.3794&$(13,24)$    &    11.1209&$(14,11)$    &     9.8142                    \\
$(14,15)$    &     9.0363&$(14,23)$    &       8.4002&$(15,10)$    &      23.1929&$(15,14)$    &       9.0798             \\
$(15,19)$    &      19.0836&$(15,22)$    &      18.4094&$(16,8 )$    &       8.4066&$(16,10)$    &      11.0728           \\
$(16,17)$    &      11.6939&$(16,18)$    &      15.2805&$(17,10)$    &       8.1000&$(17,16)$    &      11.6829           \\
$(17,19)$    &       9.9528&$(18,7 )$    &      15.8573&$(18,16)$    &      15.3354&$(18,20)$    &      18.9793           \\
$(19,15)$    &      19.1171&$(19,17)$    &       9.9417&$(19,20)$    &       8.6874&$(20,18)$    &      18.9950           \\
$(20,19)$    &       8.7098&$(20,21)$    &       6.3023&$(20,22)$    &       7.0000&$(21,20)$    &       6.2404           \\
$(21,22)$    &       8.6188&$(21,24)$    &      10.3095&$(22,15)$    &      18.3857&$(22,20)$    &       7.0000           \\
$(22,21)$    &       8.6069&$(22,23)$    &       9.6618&$(23,14)$    &       8.3945&$(23,22)$    &       9.6261           \\
$(23,24)$    &       7.9028&$(24,13)$    &      11.1122&$(24,21)$    &      10.2595&$(24,23)$    &       7.8613           \\

\bottomrule
\end{tabular*}
}
\end{table}

\begin{figure}[!ht]
\centering
\includegraphics[scale=0.6]{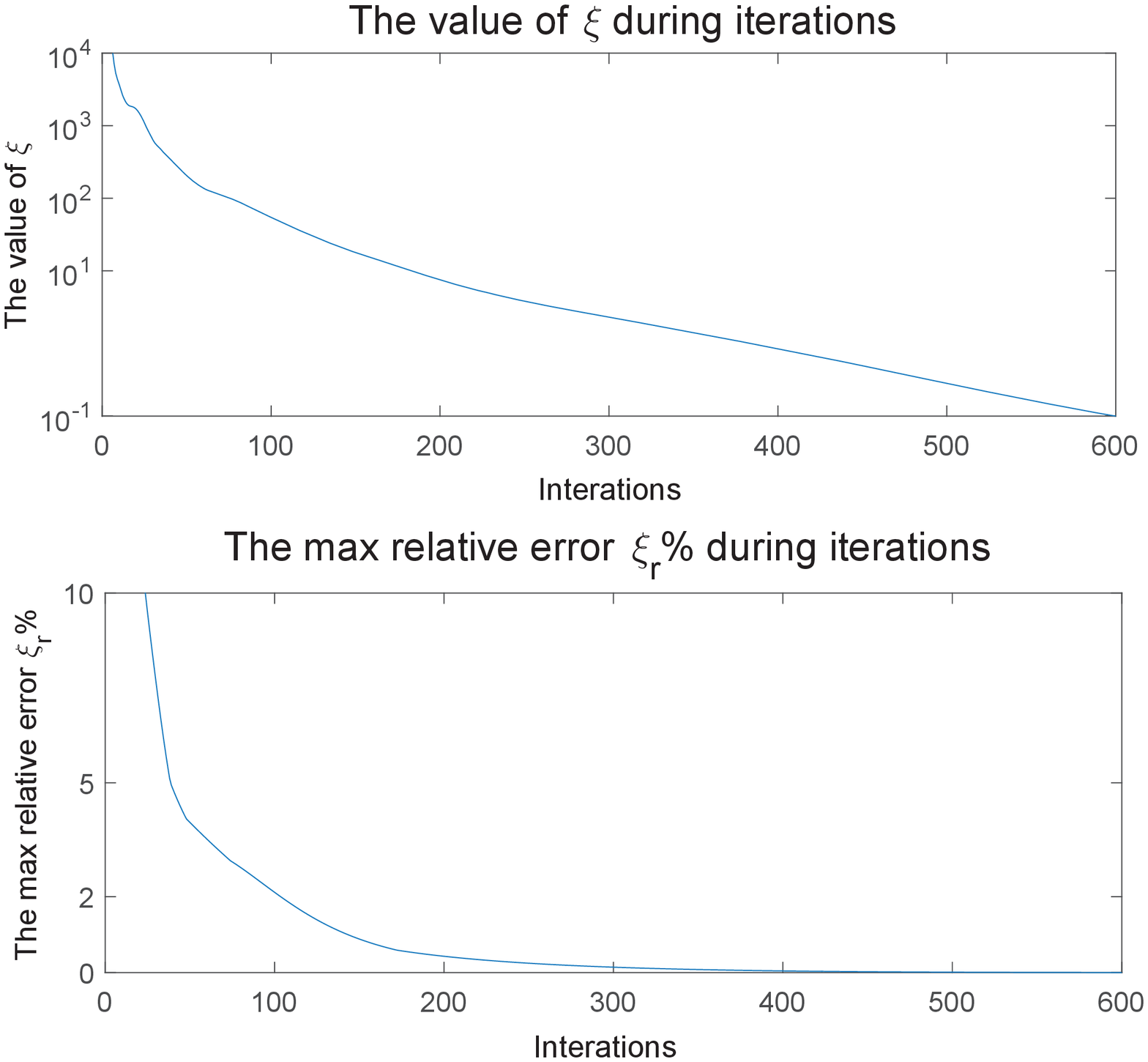}
 \caption{The error and max relative error of each link flow during iterations  } \label{ERR}
\end{figure}

Let $\varepsilon^n$ and $\varepsilon_r^n$  denote the summation of the error  and the max relative error between each  equilibrium link flow and the assignment link  flow at $n_{th}$ iteration, namely:
\begin{equation}
\begin{array}{l}
\varepsilon^n= \sum\limits_{i,j\in N}\varepsilon_{ij}^{n}, ~~~\forall i,j \in N \\
\varepsilon_r^n= \max \{\frac{{\varepsilon _{ij}^n}}{{Q_{ij}^{all}}}\}, ~~~\forall i,j \in N
\end{array}
\end{equation}
where  $Q_{ij}^{all}$ denotes the equilibrium flow along link $M_{ij}$, $\varepsilon_{ij}^{n}$ denotes the error between  equilibrium flow and the assignment flow at the $n_{th}$ iteration along link $M_{ij}$.
The change of $\varepsilon$ and $\varepsilon_r$ during iterations are indicated in Figure \ref{ERR}.
The summation of the error $\varepsilon$ and  the maximum relative error $\varepsilon_r$ are monotone decreasing during iterations.
The maximum relative error of each link flow $\varepsilon_r^{24}$ is no more than $10\%$ at the $24_{th}$ iteration. After 100 iterations, the maximum relative error of each link flow $\varepsilon_r^{100}$ is no more than $2\%$ and the summation of error between each equilibrium link flow and the assignment link  flow $\varepsilon^{100}$ equals  54.2587.

\section{CONClUSIONS}

To address UE traffic assignment problem, a modified \emph{Physarum}-inspired model for UE traffic assignment is proposed in this paper. Considering the foraging behavior of \emph{Physarum}, the equilibrium flows can be obtained when the \emph{Physarum} can't find a shorter travel time between each OD pair. The modified  algorithm is more efficient in real traffic networks with two-way traffic characteristics and  multiple OD pairs .
By decomposing flows according to origin node, the flows from different OD pairs can be distinguished. Numerical examples are illustrated to show  the rationality and convergence properties of the proposed algorithm.

 In the future, one of our works is to  use paralleled computing and the optimal model for the linear system of
equations \cite{koutis2010approaching} to reduce computing time. Besides, the theoretical analysis of convergence is also our research topic.

\section*{Acknowledgment}

The work is partially supported by National Natural Science Foundation of China (Grant No. 61671384), Natural Science Basic Research Plan in Shaanxi Province of China (Program No. 2016JM6018), the Fund of SAST (Program No. SAST2016083), the Seed Foundation of Innovation and Creation for Graduate Students in Northwestern Polytechnical University (Program No. Z2016122).

\bibliographystyle{model1-num-names}
\bibliography{Sxxreference}

\begin{thebibliography}{55}
\expandafter\ifx\csname natexlab\endcsname\relax\def\natexlab#1{#1}\fi
\providecommand{\bibinfo}[2]{#2}
\ifx\xfnm\relax \def\xfnm[#1]{\unskip,\space#1}\fi
\bibitem[{Bertsekas and Gafni(1982)}]{bertsekas1982projection}
\bibinfo{author}{D.~P. Bertsekas}, \bibinfo{author}{E.~M. Gafni},
\newblock \bibinfo{title}{Projection methods for variational inequalities with
  application to the traffic assignment problem},
\newblock in: \bibinfo{booktitle}{Nondifferential and Variational Techniques in
  Optimization}, \bibinfo{publisher}{Springer}, \bibinfo{year}{1982}, pp.
  \bibinfo{pages}{139--159}.
\bibitem[{Papageorgiou(1990)}]{papageorgiou1990dynamic}
\bibinfo{author}{M.~Papageorgiou},
\newblock \bibinfo{title}{Dynamic modeling, assignment, and route guidance in
  traffic networks},
\newblock \bibinfo{journal}{Transportation Research Part B: Methodological}
  \bibinfo{volume}{24} (\bibinfo{year}{1990}) \bibinfo{pages}{471--495}.
\bibitem[{Ziliaskopoulos(2000)}]{ISI:000085665800003}
\bibinfo{author}{A.~Ziliaskopoulos},
\newblock \bibinfo{title}{{A linear programming model for the single
  destination System Optimum Dynamic Traffic Assignment problem}},
\newblock \bibinfo{journal}{{Transportation Science}} \bibinfo{volume}{{34}}
  (\bibinfo{year}{{2000}}) \bibinfo{pages}{{37--49}}.
\bibitem[{Liu et~al.(2010)Liu, Bunker, and Ferreira}]{ISI:000283918200005}
\bibinfo{author}{Y.~Liu}, \bibinfo{author}{J.~Bunker},
  \bibinfo{author}{L.~Ferreira},
\newblock \bibinfo{title}{{Transit Users' Route-Choice Modelling in Transit
  Assignment: A Review}},
\newblock \bibinfo{journal}{{Transport Reviews}} \bibinfo{volume}{{30}}
  (\bibinfo{year}{{2010}}) \bibinfo{pages}{{753--769}}.
\bibitem[{Du et~al.(2016)Du, Zhou, Lordan, Wang, Zhao, and
  Zhu}]{du2016analysis}
\bibinfo{author}{W.-B. Du}, \bibinfo{author}{X.-L. Zhou},
  \bibinfo{author}{O.~Lordan}, \bibinfo{author}{Z.~Wang},
  \bibinfo{author}{C.~Zhao}, \bibinfo{author}{Y.-B. Zhu},
\newblock \bibinfo{title}{Analysis of the chinese airline network as
  multi-layer networks},
\newblock \bibinfo{journal}{Transportation Research Part E: Logistics and
  Transportation Review} \bibinfo{volume}{89} (\bibinfo{year}{2016})
  \bibinfo{pages}{108--116}.
\bibitem[{Wardrop(1952)}]{Wardrop1952Some}
\bibinfo{author}{J.~G. Wardrop},
\newblock \bibinfo{title}{Some theoretical aspects of road traffic research},
\newblock in: \bibinfo{booktitle}{Inst Civil Engineers Proc London /UK/}, pp.
  \bibinfo{pages}{72--73}.
\bibitem[{Beckmann et~al.(1956)Beckmann, Mcguire, Winsten, and
  Koopmans}]{Beckmann1956Studies}
\bibinfo{author}{M.~J. Beckmann}, \bibinfo{author}{C.~B. Mcguire},
  \bibinfo{author}{C.~B. Winsten}, \bibinfo{author}{T.~C. Koopmans},
\newblock \bibinfo{title}{Studies in the economics of transportation},
\newblock \bibinfo{journal}{Economic Journal} \bibinfo{volume}{26}
  (\bibinfo{year}{1956}) \bibinfo{pages}{820--821}.
\bibitem[{Chiou(2009)}]{Chiou2009An}
\bibinfo{author}{S.~W. Chiou},
\newblock \bibinfo{title}{An efficient algorithm for optimal design of area
  traffic control with network flows},
\newblock \bibinfo{journal}{Applied Mathematical Modelling}
  \bibinfo{volume}{33} (\bibinfo{year}{2009}) \bibinfo{pages}{2710--2722}.
\bibitem[{Florian et~al.(2009)Florian, Constantin, and
  Florian}]{florian2009new}
\bibinfo{author}{M.~Florian}, \bibinfo{author}{I.~Constantin},
  \bibinfo{author}{D.~Florian},
\newblock \bibinfo{title}{A new look at projected gradient method for
  equilibrium assignment},
\newblock \bibinfo{journal}{Transportation Research Record: Journal of the
  Transportation Research Board}  (\bibinfo{year}{2009})
  \bibinfo{pages}{10--16}.
\bibitem[{Chiou(2010)}]{Chiou2010An}
\bibinfo{author}{S.~W. Chiou},
\newblock \bibinfo{title}{An efficient algorithm for computing traffic
  equilibria using transyt model},
\newblock \bibinfo{journal}{Applied Mathematical Modelling}
  \bibinfo{volume}{34} (\bibinfo{year}{2010}) \bibinfo{pages}{3390--3399}.
\bibitem[{Lin and Leong(2014)}]{Lin2014An}
\bibinfo{author}{D.~Y. Lin}, \bibinfo{author}{P.~W. Leong},
\newblock \bibinfo{title}{An n-path user equilibrium for transportation
  networks},
\newblock \bibinfo{journal}{Applied Mathematical Modelling}
  \bibinfo{volume}{38} (\bibinfo{year}{2014}) \bibinfo{pages}{667¨C682}.
\bibitem[{Leblanc et~al.(1975)Leblanc, Morlok, and Pierskalla}]{Leblanc1975An}
\bibinfo{author}{L.~J. Leblanc}, \bibinfo{author}{E.~K. Morlok},
  \bibinfo{author}{W.~P. Pierskalla},
\newblock \bibinfo{title}{An efficient approach to solving the road network
  equilibrium traffic assignment problem},
\newblock \bibinfo{journal}{Transportation Research} \bibinfo{volume}{9}
  (\bibinfo{year}{1975}) \bibinfo{pages}{309--318}.
\bibitem[{Ryu et~al.(2014)Ryu, Chen, and Choi}]{ryu2014modified}
\bibinfo{author}{S.~Ryu}, \bibinfo{author}{A.~Chen}, \bibinfo{author}{K.~Choi},
\newblock \bibinfo{title}{A modified gradient projection algorithm for solving
  the elastic demand traffic assignment problem},
\newblock \bibinfo{journal}{Computers \& Operations Research}
  \bibinfo{volume}{47} (\bibinfo{year}{2014}) \bibinfo{pages}{61--71}.
\bibitem[{Larsson and Patriksson(1992)}]{larsson1992simplicial}
\bibinfo{author}{T.~Larsson}, \bibinfo{author}{M.~Patriksson},
\newblock \bibinfo{title}{Simplicial decomposition with disaggregated
  representation for the traffic assignment problem},
\newblock \bibinfo{journal}{Transportation Science} \bibinfo{volume}{26}
  (\bibinfo{year}{1992}) \bibinfo{pages}{4--17}.
\bibitem[{Jayakrishnan et~al.(1994)Jayakrishnan, Tsai, Prashker, and
  Rajadhyaksha}]{jayakrishnan1994faster}
\bibinfo{author}{R.~Jayakrishnan}, \bibinfo{author}{W.~T. Tsai},
  \bibinfo{author}{J.~N. Prashker}, \bibinfo{author}{S.~Rajadhyaksha},
\newblock \bibinfo{title}{A faster path-based algorithm for traffic
  assignment},
\newblock \bibinfo{journal}{University of California Transportation Center}
  (\bibinfo{year}{1994}).
\bibitem[{Cheng et~al.(2003)Cheng, Iida, Uno, and Wang}]{cheng2003alternative}
\bibinfo{author}{L.~Cheng}, \bibinfo{author}{Y.~Iida},
  \bibinfo{author}{N.~Uno}, \bibinfo{author}{W.~Wang},
\newblock \bibinfo{title}{Alternative quasi-newton methods for capacitated user
  equilibrium assignment},
\newblock \bibinfo{journal}{Transportation Research Record: Journal of the
  Transportation Research Board}  (\bibinfo{year}{2003})
  \bibinfo{pages}{109--116}.
\bibitem[{Sun et~al.(1996)Sun, Jayakrishnan, and Tsai}]{sun1996computational}
\bibinfo{author}{C.~Sun}, \bibinfo{author}{R.~Jayakrishnan},
  \bibinfo{author}{W.~Tsai},
\newblock \bibinfo{title}{Computational study of a path-based algorithm and its
  variants for static traffic assignment},
\newblock \bibinfo{journal}{Transportation Research Record: Journal of the
  Transportation Research Board}  (\bibinfo{year}{1996})
  \bibinfo{pages}{106--115}.
\bibitem[{Tatineni et~al.(1998)Tatineni, Edwards, and
  Boyce}]{tatineni1998comparison}
\bibinfo{author}{M.~Tatineni}, \bibinfo{author}{H.~Edwards},
  \bibinfo{author}{D.~Boyce},
\newblock \bibinfo{title}{Comparison of disaggregate simplicial decomposition
  and frank-wolfe algorithms for user-optimal route choice},
\newblock \bibinfo{journal}{Transportation Research Record: Journal of the
  Transportation Research Board}  (\bibinfo{year}{1998})
  \bibinfo{pages}{157--162}.
\bibitem[{Bar-Gera(1999)}]{bar1999origin}
\bibinfo{author}{H.~Bar-Gera},
\newblock \bibinfo{title}{Origin-based algorithms for transportation network
  modeling.},
\newblock \bibinfo{journal}{National Institute of Statistical Sciences Research
  Park Nc}  (\bibinfo{year}{1999}).
\bibitem[{Bar-Gera(2002)}]{bar2002origin}
\bibinfo{author}{H.~Bar-Gera},
\newblock \bibinfo{title}{Origin-based algorithm for the traffic assignment
  problem},
\newblock \bibinfo{journal}{Transportation Science} \bibinfo{volume}{36}
  (\bibinfo{year}{2002}) \bibinfo{pages}{398--417}.
\bibitem[{Dial(2006)}]{dial2006path}
\bibinfo{author}{R.~B. Dial},
\newblock \bibinfo{title}{A path-based user-equilibrium traffic assignment
  algorithm that obviates path storage and enumeration},
\newblock \bibinfo{journal}{Transportation Research Part B: Methodological}
  \bibinfo{volume}{40} (\bibinfo{year}{2006}) \bibinfo{pages}{917--936}.
\bibitem[{Xu et~al.(2008)Xu, Chen, and Gao}]{Xu2008An}
\bibinfo{author}{M.~Xu}, \bibinfo{author}{A.~Chen}, \bibinfo{author}{Z.~Gao},
\newblock \bibinfo{title}{An improved origin-based algorithm for solving the
  combined distribution and assignment problem},
\newblock \bibinfo{journal}{European Journal of Operational Research}
  \bibinfo{volume}{188} (\bibinfo{year}{2008}) \bibinfo{pages}{354--369}.
\bibitem[{Matteucci and Mussone(2013)}]{Matteucci2013An}
\bibinfo{author}{M.~Matteucci}, \bibinfo{author}{L.~Mussone},
\newblock \bibinfo{title}{An ant colony system for transportation user
  equilibrium analysis in congested networks},
\newblock \bibinfo{journal}{Swarm Intelligence} \bibinfo{volume}{7}
  (\bibinfo{year}{2013}) \bibinfo{pages}{255--277}.
\bibitem[{D¡¯Acierno et~al.(2012)D¡¯Acierno, Gallo, and Montella}]{d2012ant}
\bibinfo{author}{L.~D¡¯Acierno}, \bibinfo{author}{M.~Gallo},
  \bibinfo{author}{B.~Montella},
\newblock \bibinfo{title}{An ant colony optimisation algorithm for solving the
  asymmetric traffic assignment problem},
\newblock \bibinfo{journal}{European Journal of Operational Research}
  \bibinfo{volume}{217} (\bibinfo{year}{2012}) \bibinfo{pages}{459--469}.
\bibitem[{Sanchez-Medina et~al.(2012)Sanchez-Medina, Diaz-Cabrera,
  Galan-Moreno, and Rubio-Royo}]{MorenoDiaz2012}
\bibinfo{author}{J.~J. Sanchez-Medina}, \bibinfo{author}{M.~Diaz-Cabrera},
  \bibinfo{author}{M.~J. Galan-Moreno}, \bibinfo{author}{E.~Rubio-Royo},
\newblock \bibinfo{title}{{User Equilibrium Study of AETROS Travel Route
  Optimization System}},
\newblock in: \bibinfo{editor}{{MorenoDiaz, R and Pichler, F and
  QuesadaArencibia, A}} (Ed.), \bibinfo{booktitle}{{Computer Aided Systems
  Theory - Eurocast 2011, PT II}}, volume \bibinfo{volume}{{6928}} of
  \textit{\bibinfo{series}{{Lecture Notes in Computer Science}}},
  \bibinfo{publisher}{{Springer-Verlag Berlin}},
  \bibinfo{address}{{Heidelberger Platz 3, D-14197 Berlin, Germany}},
  \bibinfo{year}{{2012}}, pp. \bibinfo{pages}{{465--472}}.
\bibitem[{Adamatzky(2007)}]{adamatzky2007physarum}
\bibinfo{author}{A.~Adamatzky},
\newblock \bibinfo{title}{Physarum machines: encapsulating reaction--diffusion
  to compute spanning tree},
\newblock \bibinfo{journal}{Naturwissenschaften} \bibinfo{volume}{94}
  (\bibinfo{year}{2007}) \bibinfo{pages}{975--980}.
\bibitem[{Adamatzky and Jones(2010)}]{adamatzky2010programmable}
\bibinfo{author}{A.~Adamatzky}, \bibinfo{author}{J.~Jones},
\newblock \bibinfo{title}{Programmable reconfiguration of physarum machines},
\newblock \bibinfo{journal}{Natural Computing} \bibinfo{volume}{9}
  (\bibinfo{year}{2010}) \bibinfo{pages}{219--237}.
\bibitem[{Adamatzky and Jones(2015)}]{adamatzky2015using}
\bibinfo{author}{A.~Adamatzky}, \bibinfo{author}{J.~Jones},
\newblock \bibinfo{title}{On using compressibility to detect when slime mould
  completed computation},
\newblock \bibinfo{journal}{Complexity}  (\bibinfo{year}{2015}).
\bibitem[{Stephenson and Stempen(1995)}]{Stephenson1995Myxomycetes}
\bibinfo{author}{S.~L. Stephenson}, \bibinfo{author}{H.~Stempen},
\newblock \bibinfo{title}{Myxomycetes : a handbook of slime molds},
\newblock \bibinfo{journal}{Bioscience} \bibinfo{volume}{45}
  (\bibinfo{year}{1995}) \bibinfo{pages}{601--602}.
\bibitem[{Nakagaki et~al.(2000)Nakagaki, Yamada, and
  T{\'o}th}]{nakagaki2000intelligence}
\bibinfo{author}{T.~Nakagaki}, \bibinfo{author}{H.~Yamada},
  \bibinfo{author}{{\'A}.~T{\'o}th},
\newblock \bibinfo{title}{Intelligence: Maze-solving by an amoeboid organism},
\newblock \bibinfo{journal}{Nature} \bibinfo{volume}{407}
  (\bibinfo{year}{2000}) \bibinfo{pages}{470--470}.
\bibitem[{Tero et~al.(2010)Tero, Takagi, Saigusa, Ito, Bebber, Fricker, Yumiki,
  Kobayashi, and Nakagaki}]{Tero2010Rules}
\bibinfo{author}{A.~Tero}, \bibinfo{author}{S.~Takagi},
  \bibinfo{author}{T.~Saigusa}, \bibinfo{author}{K.~Ito},
  \bibinfo{author}{D.~P. Bebber}, \bibinfo{author}{M.~D. Fricker},
  \bibinfo{author}{K.~Yumiki}, \bibinfo{author}{R.~Kobayashi},
  \bibinfo{author}{T.~Nakagaki},
\newblock \bibinfo{title}{Rules for biologically inspired adaptive network
  design.},
\newblock \bibinfo{journal}{Science} \bibinfo{volume}{327}
  (\bibinfo{year}{2010}) \bibinfo{pages}{439--442}.
\bibitem[{Bonifaci et~al.(2012)Bonifaci, Mehlhorn, and
  Varma}]{Bonifaci2012Physarum}
\bibinfo{author}{V.~Bonifaci}, \bibinfo{author}{K.~Mehlhorn},
  \bibinfo{author}{G.~Varma},
\newblock \bibinfo{title}{Physarum can compute shortest paths},
\newblock in: \bibinfo{booktitle}{Acm-Siam Symposium on Discrete Algorithms},
  p. \bibinfo{pages}{121¨C133}.
\bibitem[{Adamatzky(2012)}]{Adamatzky2012Slime}
\bibinfo{author}{A.~Adamatzky},
\newblock \bibinfo{title}{Slime mold solves maze in one pass, assisted by
  gradient of chemo-attractants.},
\newblock \bibinfo{journal}{IEEE Transactions on Nanobioscience}
  \bibinfo{volume}{11} (\bibinfo{year}{2012}) \bibinfo{pages}{131--4}.
\bibitem[{Wang et~al.(2014)Wang, Lu, Zhang, Wang, and Deng}]{Wang2014A}
\bibinfo{author}{H.~Wang}, \bibinfo{author}{X.~Lu}, \bibinfo{author}{X.~Zhang},
  \bibinfo{author}{Q.~Wang}, \bibinfo{author}{Y.~Deng},
\newblock \bibinfo{title}{A bio-inspired method for the constrained shortest
  path problem},
\newblock \bibinfo{journal}{The Scientific World Journal}
  \bibinfo{volume}{2014} (\bibinfo{year}{2014}).
\bibitem[{Zhang et~al.(2014)Zhang, Zhang, and Deng}]{Zhang2014An}
\bibinfo{author}{X.~Zhang}, \bibinfo{author}{Y.~Zhang},
  \bibinfo{author}{Y.~Deng},
\newblock \bibinfo{title}{{An improved bio-inspired algorithm for the directed
  shortest path problem}},
\newblock \bibinfo{journal}{Bioinspiration \& Biomimetics}
  \bibinfo{volume}{{9}} (\bibinfo{year}{{2014}}).
\bibitem[{Wang et~al.(2015)Wang, Lu, Zhang, Deng, and Xiao}]{Wang2015An}
\bibinfo{author}{Q.~Wang}, \bibinfo{author}{X.~Lu}, \bibinfo{author}{X.~Zhang},
  \bibinfo{author}{Y.~Deng}, \bibinfo{author}{C.~Xiao},
\newblock \bibinfo{title}{{An anticipation mechanism for the shortest path
  problem based on Physarum polycephalum}},
\newblock \bibinfo{journal}{{International Journal of General Systems}}
  \bibinfo{volume}{{44}} (\bibinfo{year}{{2015}}) \bibinfo{pages}{{326--340}}.
\bibitem[{Adamatzky(2012)}]{Adamatzky2012Bioevaluation}
\bibinfo{author}{A.~Adamatzky},
\newblock \bibinfo{title}{Bioevaluation of world transport networks},
\newblock \bibinfo{journal}{Bioevaluation of World Transport Networks}
  \bibinfo{volume}{43} (\bibinfo{year}{2012}) \bibinfo{pages}{368}.
\bibitem[{Adamatzky et~al.(2013)Adamatzky, Lees, and Sloot}]{ANDREW2013BIO}
\bibinfo{author}{A.~Adamatzky}, \bibinfo{author}{M.~Lees},
  \bibinfo{author}{P.~Sloot},
\newblock \bibinfo{title}{Bio-development of motorway network in the
  netherlands: a slime mould approach},
\newblock \bibinfo{journal}{Advances in Complex Systems} \bibinfo{volume}{16}
  (\bibinfo{year}{2013}) \bibinfo{pages}{1250034}.
\bibitem[{Tsompanas et~al.(2014)Tsompanas, Sirakoulis, and
  Adamatzky}]{tsompanas2014physarum}
\bibinfo{author}{M.-A.~I. Tsompanas}, \bibinfo{author}{G.~C. Sirakoulis},
  \bibinfo{author}{A.~I. Adamatzky},
\newblock \bibinfo{title}{Physarum in silicon: the greek motorways study},
\newblock \bibinfo{journal}{Natural Computing}  (\bibinfo{year}{2014})
  \bibinfo{pages}{1--17}.
\bibitem[{Evangelidis et~al.(2015)Evangelidis, Tsompanas, Sirakoulis, and
  Adamatzky}]{Evangelidis2015Slime}
\bibinfo{author}{V.~Evangelidis}, \bibinfo{author}{M.~A. Tsompanas},
  \bibinfo{author}{G.~C. Sirakoulis}, \bibinfo{author}{A.~Adamatzky},
\newblock \bibinfo{title}{Slime mould imitates development of roman roads in
  the balkans},
\newblock \bibinfo{journal}{Journal of Archaeological Science Reports}
  \bibinfo{volume}{2} (\bibinfo{year}{2015}) \bibinfo{pages}{264--281}.
\bibitem[{Schumann and Adamatzky(2011{\natexlab{a}})}]{schumann2011logical}
\bibinfo{author}{A.~Schumann}, \bibinfo{author}{A.~Adamatzky},
\newblock \bibinfo{title}{Logical modelling of physarum polycephalum},
\newblock \bibinfo{journal}{arXiv preprint arXiv:1105.4060}
  (\bibinfo{year}{2011}{\natexlab{a}}).
\bibitem[{Schumann and Adamatzky(2011{\natexlab{b}})}]{schumann2011physarum}
\bibinfo{author}{A.~Schumann}, \bibinfo{author}{A.~Adamatzky},
\newblock \bibinfo{title}{Physarum spatial logic},
\newblock \bibinfo{journal}{New Mathematics and Natural Computation}
  \bibinfo{volume}{7} (\bibinfo{year}{2011}{\natexlab{b}})
  \bibinfo{pages}{483--498}.
\bibitem[{Braund and Miranda(2013)}]{braund2013music}
\bibinfo{author}{E.~Braund}, \bibinfo{author}{E.~Miranda},
\newblock \bibinfo{title}{Music with unconventional computing: a system for
  physarum polycephalum sound synthesis},
\newblock in: \bibinfo{booktitle}{International Symposium on Computer Music
  Modeling and Retrieval}, \bibinfo{organization}{Springer}, pp.
  \bibinfo{pages}{175--189}.
\bibitem[{Miranda(2014)}]{miranda2014harnessing}
\bibinfo{author}{E.~R. Miranda},
\newblock \bibinfo{title}{Harnessing the intelligence of physarum polycephalum
  for unconventional computing-aided musical composition.},
\newblock \bibinfo{journal}{IJUC} \bibinfo{volume}{10} (\bibinfo{year}{2014})
  \bibinfo{pages}{251--268}.
\bibitem[{Braund and Miranda(2015)}]{braund2015music}
\bibinfo{author}{E.~Braund}, \bibinfo{author}{E.~Miranda},
\newblock \bibinfo{title}{Music with unconventional computing: towards a step
  sequencer from plasmodium of physarum polycephalum},
\newblock in: \bibinfo{booktitle}{International Conference on Evolutionary and
  Biologically Inspired Music and Art}, \bibinfo{organization}{Springer}, pp.
  \bibinfo{pages}{15--26}.
\bibitem[{Zhang et~al.(2013)Zhang, Zhang, Deng, and
  Mahadevan}]{Zhangya2013fuzzy}
\bibinfo{author}{Y.~Zhang}, \bibinfo{author}{Z.~Zhang},
  \bibinfo{author}{Y.~Deng}, \bibinfo{author}{S.~Mahadevan},
\newblock \bibinfo{title}{{A biologically inspired solution for fuzzy shortest
  path problems}},
\newblock \bibinfo{journal}{Applied Soft Computing} \bibinfo{volume}{{13}}
  (\bibinfo{year}{{2013}}) \bibinfo{pages}{{2356--2363}}.
\bibitem[{Zhang et~al.(2014)Zhang, Wang, Adamatzky, Chan, Mahadevan, and
  Deng}]{Zhang2014fuzzy}
\bibinfo{author}{X.~Zhang}, \bibinfo{author}{Q.~Wang},
  \bibinfo{author}{A.~Adamatzky}, \bibinfo{author}{F.~T. Chan},
  \bibinfo{author}{S.~Mahadevan}, \bibinfo{author}{Y.~Deng},
\newblock \bibinfo{title}{A biologically inspired optimization algorithm for
  solving fuzzy shortest path problems with mixed fuzzy arc lengths},
\newblock \bibinfo{journal}{Journal of Optimization Theory \& Applications}
  \bibinfo{volume}{163} (\bibinfo{year}{2014}) \bibinfo{pages}{1049--1056}.
\bibitem[{Jiang et~al.(2015)Jiang, Luo, Qin, and Zhan}]{Jiang2015improved}
\bibinfo{author}{W.~Jiang}, \bibinfo{author}{Y.~Luo}, \bibinfo{author}{X.~Qin},
  \bibinfo{author}{J.~Zhan},
\newblock \bibinfo{title}{An improved method to rank generalized fuzzy numbers
  with different left heights and right heights},
\newblock \bibinfo{journal}{Journal of Intelligent \& Fuzzy Systems}
  \bibinfo{volume}{28} (\bibinfo{year}{2015}) \bibinfo{pages}{2343--2355}.
\bibitem[{Zhang et~al.(2015)Zhang, Adamatzky, Chan, Deng, Yang, Yang,
  Tsompanas, Sirakoulis, and Mahadevan}]{Zhang2015A}
\bibinfo{author}{X.~Zhang}, \bibinfo{author}{A.~Adamatzky},
  \bibinfo{author}{F.~T. Chan}, \bibinfo{author}{Y.~Deng},
  \bibinfo{author}{H.~Yang}, \bibinfo{author}{X.-S. Yang},
  \bibinfo{author}{M.-A.~I. Tsompanas}, \bibinfo{author}{G.~C. Sirakoulis},
  \bibinfo{author}{S.~Mahadevan},
\newblock \bibinfo{title}{A biologically inspired network design model},
\newblock \bibinfo{journal}{Scientific reports} \bibinfo{volume}{5}
  (\bibinfo{year}{2015}).
\bibitem[{Zhang(2015)}]{Zhang2015AnEfficient}
\bibinfo{author}{X.~Zhang},
\newblock \bibinfo{title}{{An Efficient Physarum Algorithm for Solving the
  Bicriteria Traffic Assignment Problem}},
\newblock \bibinfo{journal}{{International Journal of Unconventional
  Computing}} \bibinfo{volume}{{11}} (\bibinfo{year}{{2015}})
  \bibinfo{pages}{{473--490}}.
\bibitem[{Tero et~al.(2007)Tero, Kobayashi, and
  Nakagaki}]{tero2007mathematical}
\bibinfo{author}{A.~Tero}, \bibinfo{author}{R.~Kobayashi},
  \bibinfo{author}{T.~Nakagaki},
\newblock \bibinfo{title}{A mathematical model for adaptive transport network
  in path finding by true slime mold},
\newblock \bibinfo{journal}{Journal of theoretical biology}
  \bibinfo{volume}{244} (\bibinfo{year}{2007}) \bibinfo{pages}{553--564}.
\bibitem[{Hansen(1959)}]{hansen1959accessibility}
\bibinfo{author}{W.~G. Hansen},
\newblock \bibinfo{title}{How accessibility shapes land use},
\newblock \bibinfo{journal}{Journal of the American Institute of planners}
  \bibinfo{volume}{25} (\bibinfo{year}{1959}) \bibinfo{pages}{73--76}.
\bibitem[{Boyce et~al.(2004)Boyce, Ralevic-Dekic, and
  Bar-Gera}]{boyce2004convergence}
\bibinfo{author}{D.~Boyce}, \bibinfo{author}{B.~Ralevic-Dekic},
  \bibinfo{author}{H.~Bar-Gera},
\newblock \bibinfo{title}{Convergence of traffic assignments: how much is
  enough?},
\newblock \bibinfo{journal}{Journal of Transportation Engineering}
  \bibinfo{volume}{130} (\bibinfo{year}{2004}) \bibinfo{pages}{49--55}.
\bibitem[{Adamatzky and Jones(2008)}]{adamatzky2008towards}
\bibinfo{author}{A.~Adamatzky}, \bibinfo{author}{J.~Jones},
\newblock \bibinfo{title}{Towards physarum robots: computing and manipulating
  on water surface},
\newblock \bibinfo{journal}{Journal of Bionic Engineering} \bibinfo{volume}{5}
  (\bibinfo{year}{2008}) \bibinfo{pages}{348--357}.
\bibitem[{Koutis et~al.(2010)Koutis, Miller, and Peng}]{koutis2010approaching}
\bibinfo{author}{I.~Koutis}, \bibinfo{author}{G.~L. Miller},
  \bibinfo{author}{R.~Peng},
\newblock \bibinfo{title}{Approaching optimality for solving sdd linear
  systems},
\newblock in: \bibinfo{booktitle}{Foundations of Computer Science (FOCS), 2010
  51st Annual IEEE Symposium on}, \bibinfo{organization}{IEEE}, pp.
  \bibinfo{pages}{235--244}.

\end{thebibliography}
\end{document}